\documentclass[a4paper,11pt]{article}
\usepackage{amssymb,palatino}
\newtheorem{theorem}{Theorem}

\newtheorem{definition}{Definition}
\newtheorem{proposition}{Proposition} 

\newcommand{\QED}{\begin{flushright} $\Box$ \end{flushright}}
\newcommand{\jacobi}[2]{\left( \begin{array}{c} #1\\ \hline #2 \end{array} \right)}

\begin{document}

\title{On the Rabin signature\thanks{A preliminary version of this paper was presented at Workshop on Computational Security, Centre de Recerca Matem\`atica (CRM), Bellaterra (Barcelona), 28 November-02 December 2011.}}

\author{Michele Elia\thanks{Politecnico di Torino, Italy}, ~~
 Davide Schipani \thanks{University of Zurich, Switzerland}
}

\maketitle

\thispagestyle{empty}

\begin{abstract}
\noindent
Some Rabin signature schemes may be exposed to forgery; several variants are here
 described to counter this vulnerability. 
Blind Rabin signatures are also discussed.
\end{abstract}

\paragraph{Keywords:} Rabin signature, Rabin cryptosystem.

\vspace{2mm}
\noindent
{\bf Mathematics Subject Classification (2010): 94A60, 11T71, 14G50}

\vspace{8mm}

\section{Introduction}
The public-key cryptosystems based on the Rabin scheme have in principle
 two main advantages with respect to some other alternative public-key schemes,
 namely, they are provably as hard to break as factoring, and they should involve a smaller
 computational burden, even though practical implementations require some adjustments that diminish
 the theoretical advantages \cite{buchmann,silver,menezes3}. 
The Rabin scheme can be used in different applications, e.g. to exchange secret messages, and
 to provide electronic signatures. 
In \cite{rabin1}, the Rabin scheme was revisited mainly referring to the exchange of
 secret messages with respect to the problem of the unique identification of the root 
 at the decryption stage. 
Further a deterministic way was presented to compute the padding factor in 
 the classical Rabin signature (cf. \cite{menezes3}). However, this signature is plainly
 vulnerable to forgery attacks, a weakness that is absent in the Rabin-Williams signature (cf. \cite{galbraith,williams}). 
A blind Rabin-Williams signature was proposed in \cite{qui}, however some weaknesses
 of this signature were shown in \cite{chun}. \\
In this paper, we propose some variants of Rabin signatures and blind Rabin signatures, and discuss their resistances to forgery.

\section{Preliminaries} 

All operations are hereafter done in $\mathbb Z_N$, the residue ring modulo $N=pq$, a
 product of two primes $p$ and $q$ known only by the signer.
A valid signature of a message $m \in \mathbb Z_N$ consists of an $(\ell+1)$-tuple of elements
$  [ m, f_1,f_2, \ldots,f_{\ell}]  $ of  $\mathbb Z_N$,
together with a verifying function $\mathfrak v$ from $\mathbb Z_N^{\ell+1}$ into
 $\mathbb Z_N^k$, $k\geq 1$,
such that $\mathfrak v(m, f_1,f_2, \ldots,f_{\ell})=\mathbf 0$. More generally, the message $m$ may belong to some $\mathbb Z_M$, with $M\geq N$, but the verifying function uses $H(m)$ as input instead of $m$, where $H(.)$ is a hash function with values in $\mathbb Z_N$.
 
\vspace{3mm}

The classic Rabin signature of a message $m$  is a triple $(m,U,S)$, where $U$ is a padding factor (found either randomly \cite{seberry} or deterministically as in \cite{rabin1}) such that the equation $x^2 =mU $ is solvable, and $S$ is one of its roots.
Verification is performed by comparing $mU $ with $S^2$. 
An easy forgery attack computes $S^2$ or $mU$, chooses any message $m'$, computes
 $U'=S^2 m'^{-1} $, and forges the signature as $(m',U',S)$ without knowing the
 factorization of $N$. In the original proposal \cite{rabin}, a hash function $H(.)$ 
 is used instead of $m$, 
 and $S$ is a solution of $x^2 = H(mU) $, but this does not help
 against the above forgery attack. 
 
\vspace{3mm}

The Rabin-Williams signature (cf. \cite{galbraith,williams}), which is limited to pair of
 primes, where one is congruent to $3$ and the other to $7$ modulo $8$, avoids the forgery
 vulnerability.
The signature is a four-tuple $[m,e,f,S]$, where $e \in \{1,-1\}$ and $f \in \{1,2\}$ are chosen to make the equation $efS^2=H(m)$ solvable, where $H(.)$ is a convenient hash function.
The non-forgeability is based on the limited set of multipliers $e$ and $f$.
However, the Rabin-Williams scheme requires the use of two primes respectively congruent to $3$ and $7$ modulo $8$, while the classic Rabin signature works with every pair of primes. A possible  Rabin signature that avoids forgery and works for every pair of primes
 was devised in \cite{kurosawa}. 
 
\vspace{3mm}
Blind signature schemes are cryptographic primitives, which are useful in protocols that guarantee the anonymity of the parties. 
 They are playing an important role for e-commerce, e-money and e-voting procedures. In fact they were introduced by Chaum \cite{chaum} for privacy-related protocols where the signer and message author are different parties.
The blind signature is a form of digital signature in which a message is disguised 
 before it is signed, while the resulting signature can be publicly verified against the
 original message in the manner of a regular digital signature.
Formally, a message $m$ is disguised by means of a function $\mathfrak d$
 and then submitted to the signer. The signed message $[\mathfrak d(m), f_1,f_2, \ldots,f_{\ell}]$
 is then made public by the message author in the form of a valid signed message as
 $[m, f_1',f_2', \ldots,f_{\ell}']$. 

In our formal discussion, we need a precise notion of forgeability, and we will
 adopt the following definitions:
 
  \begin{definition}
   \label{defnonforg3}
 A signature of a message $m$, of the form $[m,f_1,f_2, \ldots,f_{\ell}]$, 
 is said to be strongly forgeable if it is feasible for an outsider to derive from it a
 valid signature $[m',f_1',f_2', \ldots,f_{\ell}']$ for a given message $m'$.  
 \end{definition}
 \begin{definition}
   \label{defnonforg4}
 A signature of a message $m$, of the form $[m,f_1,f_2, \ldots,f_{\ell}]$, 
 is said to be weakly forgeable if  it is feasible for an outsider to derive from it a
 valid signature $[m',f_1',f_2', \ldots,f_{\ell}']$ for some message $m'$. 
 \end{definition}
 \begin{definition}
   \label{defnonforg}
A signature of a message $m$, of the form $[m,f_1,f_2, \ldots,f_{\ell}]$, 
 is said to be weakly non-forgeable if it is not feasible for an outsider to derive from it a
 valid signature $[m',f_1',f_2', \ldots,f_{\ell}']$ for a given message $m'$. 
 \end{definition}
 \begin{definition}
   \label{defnonforg2} 
A signature of a message $m$, of the form $[m,f_1,f_2, \ldots,f_{\ell}]$, 
 is said to be strongly non-forgeable if it is not feasible for an outsider to derive from it a
 valid signature $[m',f_1',f_2', \ldots,f_{\ell}']$ for some message $m'$.  
\end{definition}

\noindent
In other terms, a Rabin signature $[m,f_1,f_2, \ldots,f_{\ell}]$ is strongly non-forgeable
 if we cannot derive, without knowing the factorization of $N$, a whatsoever valid
 signature $[\bar m,\bar f_1,\bar f_2, \ldots,\bar f_{\ell}]$.\\
Instead, a Rabin signature $[m,f_1,f_2, \ldots,f_{\ell}]$ is weakly non-forgeable
 if we cannot derive, without knowing the factorization of $N$, a valid signature 
 for a well specified message $m'$.

For example, the Rabin-Williams signature $[m,e,f,S]$ is weakly forgeable if the hash function is the identity function, i.e. $H(u)=u$, because we can derive a valid signature as
$[r^2m,e,f,rS]$ for every factor $r$. But, depending on the hash function, this signature may be strongly non-forgeable.
In the same way the RSA signature $[m, m^D]$, where $D$ is the secret counterpart of the
 public key $E$, is weakly forgeable because we can obtain a valid signature as $[r^Em, rm^D]$, for every
 factor $r$. 
 
These examples are instances of the following general result.

\begin{definition}
   \label{defnonforg5}
 A signature of a message $m$ is said to be pseudo-homogeneous if  there are nonnegative integers  $n_0,\ldots,n_l,t_1,\ldots,t_k$ such that each component $\mathfrak v^{i}$ of the verifying function $\mathfrak v$ satisfies 
 $$\mathfrak v^{i}(\lambda^{n_0}m, \lambda^{n_1}f_1,\lambda^{n_2}f_2, \ldots,\lambda^{n_{\ell}}f_{\ell})=\lambda^{t_i}\mathfrak v^{i}(m, f_1,f_2, \ldots,f_{\ell})~~
  ~~\forall ~\lambda\in\mathbb Z_N^*~~. $$
 In particular if $\mathfrak v$ is homogeneous of degree $t$, the signature is pseudo-homogeneous with $n_0=\ldots=n_l=1$.
 \end{definition}

\begin{proposition}
  \label{homogeneous}
A pseudo-homogeneous signature is weakly forgeable.
\end{proposition}

\noindent
{\sc Proof}. 
By definition of pseudo-homogeneity, given a valid signature $[m, f_1,f_2, \ldots,f_{\ell}]$ (therefore $\mathfrak v(m, f_1,f_2, \ldots,f_{\ell})=\mathbf 0$), the signature $[\lambda^{n_0}m, \lambda^{n_1}f_1,\lambda^{n_2}f_2, \ldots,\lambda^{n_{\ell}}f_{\ell})]$ is valid for any $\lambda\in\mathbb Z_N^*$.

\QED

In the case of blind signatures, we must
 be able to derive a valid signature $[m, f_1',f_2', \ldots,f_{\ell}']$ from the 
 signature of the blind message $[\mathfrak d(m), f_1,f_2, \ldots,f_{\ell}]$;
 as a direct consequence of the above definitions this entails the following 

\begin{proposition}
  \label{blind}
A blind signer cannot employ a strongly non-forgeable signature scheme, although 
 the signature of the unblinded message may be strongly non-forgeable.
\end{proposition}

\noindent
{\sc Proof}. The first part of the statement is a simple consequence of the fact that
 strong non-forgeability implies by definition that it is not possible to derive any other valid signature, which on the other hand must occur as a purpose of the blinding technique.  
The second part is proved by an actual instance.
Let $m$ be the message that we want to be blindly signed, then the message
 $\mathfrak d(H(m))$ is submitted to the signer, who returns $[\mathfrak d(H(m)), f_1,f_2, \ldots,f_{\ell}]$. This is
  unblinded as $[H(m), f_1',f_2', \ldots,f_{\ell}']$, but it will be used
  as $[m, f_1',f_2', \ldots,f_{\ell}']$ with the assumption that the verification
  operations should consider the hashed message. If $H(.)$ is a convenient hashed function, this signature can be strongly non-forgeable, as we see later.
\QED

\section{Schemes} 
In this section we propose a general scheme that avoids forgery, and includes the Rabin-Williams signature as a special case. Further, in the case of Blum primes, we present some other forgery resistant schemes that are based on different principles. 
In the next section,  the use of these schemes to realize blind signatures will be analyzed
 with respect to forgery. Their resistance to the so called \emph{RSA blinding attack} will also be considered. 
In view of Proposition \ref{blind}, both strongly and weakly non-forgeable signatures may be of interest for different purposes.

\subsection{A general scheme}
The following is a general scheme that works for every pair
 of primes.
 
In $\mathbb Z_N^*$ a set $\mathfrak U$ can be defined with the property that,
 for any given $z \in \mathbb Z_N^*$, there exists a multiplier $\mathfrak u\in \mathfrak U$
 which makes the equation $x^2=\mathfrak u z$ solvable.  
 In fact, it is sufficient to find $4$ numbers $a_1$,$a_2$,$b_1$,$b_2$, such that
$$ \jacobi{a_1}{p}=1, \jacobi{a_2}{p}=-1, \jacobi{b_1}{q}=1, ~\mbox{and}~
 \jacobi{b_2}{q}=-1,$$
 and form the set
$$ \mathfrak U = \{ r_1^2(a_1 \psi_1 +b_1\psi_2),~ r_2^2(a_1 \psi_1 +b_2\psi_2),~ r_3^2(a_2 \psi_1 +b_1\psi_2),~ r_4^2(a_2 \psi_1 +b_2\psi_2) \}~~,  $$
 where $r_1,r_2,r_3$, and $r_4$ are four random different numbers in
 $\mathbb Z_N^*$  (necessary to prevent an easy factorization of $N$),
and $\psi_1$ and $\psi_2$ are integers determined by the extended Euclidean algorithm
 that satisfy
$$
  \psi_1+\psi_2=1 \bmod N, \hspace{5mm} \psi_1= 0 \bmod q, \hspace{5mm} \psi_2= 0 \bmod p.
$$
Given the properties above, and writing $z$ as $z_1 \psi_1 +z_2 \psi_2$ using the Chinese Remainder Theorem, one can easily find the suitable padding factor 
 $\mathfrak u \in\mathfrak U$ such that the two conditions 
 $\jacobi{\mathfrak u z}{p}=\jacobi{\mathfrak u z}{q}=1$ are
 contemporarily satisfied. 

\vspace{3mm}
\noindent
For a Rabin-type signature the public key of each user can then consist of the triple $[N,\mathfrak U, H(.)]$, where $H(.)$ is a suitable hash function, possibly the identity function. \\
The signature process is the following

\begin{description}
  \item[Public-key:] $N$, $\mathfrak U=\{\mathfrak u_1,\mathfrak u_2, \mathfrak u_3, \mathfrak u_4 \}$, and $ H(.)$. 
  \item[Signed message:] $[m,\mathfrak u, S]$, where 
    $\mathfrak u$ is the padding factor in $\mathfrak U$ which makes
    the equation $x^2=H(m) \mathfrak u$ solvable,
    and $S$ is any solution of this equation.
  \item[Verification:] Check that $\mathfrak u$ belongs to $\mathfrak U$;  
 compute $H(m) \mathfrak u$ and $S^2$; the signature is valid if and only if
 these two numbers are equal.
\end{description}  

\noindent 
The verification cost is one square and one product in $\mathbb Z_N$, plus the evaluation cost of the hash function. \\
The main advantages of this signature with respect to forgery
 are shown in the following theorem.

\begin{theorem}
   \label{forg}
The signature $[m,\mathfrak u, S 
]$ is weakly non-forgeable. It is weakly forgeable if the hash function $H(.)$
 is the identity function $H(z)=z$, while it is strongly non-forgeable if $H(.)$ is a
 convenient hash function,
 in particular, if $H(z)=z(z+1)$ (in this case the hash function is as hard to invert as factoring, and no hardness of other problems is used).
\end{theorem}

\noindent
{\sc Proof}.    
In the relation $S^2=H(m) \mathfrak u$ the number of available padding factors
 is restricted to $4$, thus for a given $S$ and correspondingly $S^2$ only $4$ values for $H(m)$ are allowed. 
 The small number of possible padding factors is what makes the signature resistant to forgery. Precisely this implies that the signature is at least weakly non-forgeable, since it is not possible to choose any $m'$ and derive a valid signature on it.
  Furthermore, the random factors $r_i$ introduced in building $\mathfrak U$ prevent
  a factorization of $N$. This can be checked, at the creation of the public key, by
  verifying that the $\mathfrak u_i$ are not among the square roots of unity and that the differences $\mathfrak u_i- \mathfrak u_j$, with $i \neq j$,
  have no factors in common with $N$. \\
 If $H(z)=z$, the signature $[m,\mathfrak u, S]$ is pseudo-homogeneous and weakly forgeable as 
  $[r^2 m,\mathfrak u, rS]$ for any $r \in \mathbb Z_N^*$, since we have
$$  r^2m \mathfrak u = r^2 S^2  \Leftrightarrow m \mathfrak u =  S^2 ~~, $$
which is true by definition. \\
If $H(.)$ is a convenient hash function, finding $m'$ from a new $S'$ is infeasible.
The special case $H(z)=z(z+1)$ is chosen as to rely on the hardness of factoring and such that it does not make the signature pseudo-homogeneous.

\QED

\subsection{Blum primes}
If the Rabin scheme is restricted to Blum primes, then it is possible to avoid the use
 of the set of multipliers $\mathfrak U$ in at least two ways.

In Variant I, the cost to pay is a further parameter in the 
 signature, which consists of a four-tuple $[m,U,S,T]$.

Let $H(m)$ be written in the form $H(m)=m_1 \psi_1+m_2\psi_2$, with $m_1=H(m) \bmod p$ and 
$m_2=H(m) \bmod q$.  
The padding factor $U$ can be chosen deterministically as in \cite{rabin1} as
 $U= R^2\left[f_1\psi_1+f_2\psi_2\right]$, where $R$ a is random number, 
 $f_1 =  \jacobi{m_1}{p}$ and $f_2=  \jacobi{m_2}{q}$. 
In fact, the equation
$$   x^2 =H(m) U = (m_1 \psi_1 +m_2 \psi_2) (f_1 \psi_1 +f_2\psi_2) =
     m_1f_1 \psi_1 +m_2 f_2\psi_2$$
is always solvable modulo $N$, because $m_1f_1$ and $m_2 f_2$
 are clearly quadratic residues modulo $p$ and modulo $q$, respectively, since $\jacobi{m_1}{p}=\jacobi{f_1}{p}$, $\jacobi{m_2}{q}=\jacobi{f_2}{q}$, so that
$$   \jacobi{m_1 f_1}{p} = \jacobi{m_1}{p}\jacobi{f_1}{p} = 1 ~,  ~ \jacobi{m_2 f_2}{q} = \jacobi{m_2}{q}\jacobi{f_2}{q} = 1  ~~. $$

Then $S$ is chosen among the roots of the equation $x^2=H(m)U$ with the further constraint that the equation $y^2=(U+1)S$ is solvable.
This is always possible because in the case of Blum primes the four roots of a quadratic
 equation form a complete set 
 of padding factors as above. \\
Lastly, $T$ is a root of $y^2=(U+1)S$.

 In Variant II, the padding factor is a square root of unity, but it is not a public element of the signature. In this case a triple will be sufficient to define a signature that is resistant to forgery.
 
\paragraph{Variant I}. \\ 
The signature process is the following:  
\begin{description}
  \item[Public-key:] $[N,H(.)]$ 
  \item[Signed message:] $[m,U,S,T]$, where 
    $U$ is a padding factor which makes the equation $x^2=H(m) U$ solvable,
    and $S$ is a root of this equation such that the equation $y^2=(U+1)S$ is solvable,
    then $T$ is any root of this equation.
  \item[Verification:] Check whether $T^2=(U+1)S$, then check whether $S^2=H(m)U$;  
    the signature is valid if and only if both equalities hold.
\end{description}  

\noindent 
The verification cost is two squares and two products in $\mathbb Z_N$, plus the evaluation of a hash function.
Note that, if $U$ is chosen deterministically as above, it is possible to make different signatures of the same message. Clearly, $U$ should not be $\psi_1-\psi_2$ or $-\psi_1+\psi_2$, because these square roots of unity would unveil the factorization of $N$; in fact adding $1$ to either of them gives a multiple of $p$ or a multiple of $q$.
Lastly, the signature is forgery resistant 
as proved in the following theorem.

\begin{theorem}
   \label{forgV1}
The signature $[m,U, S, T]$ is weakly non-forgeable. 
It is weakly forgeable if $H(z)=z$ and strongly non-forgeable
 if $H(.)$ is a convenient hash function,
 in particular, if $H(z)=z(z+1)$.
\end{theorem}

\noindent
{\sc Proof}.    
A forged signature for a given message $m'$ 
has to involve a new $U'$ and possibly a new $S'$. In either case finding the new $T'$, root of a second degree equation, requires the knowledge of the factorization of $N$. Therefore the signature is weakly non-forgeable.

If $H(z)=z$ the signature is weakly forgeable, by taking a new $T'$, finding suitable $S'$ and $U'$ and finally $m'=S'^2/U'$. If $H(.)$ is a convenient hash function, in particular,
 if $H(z)=z(z+1)$, finding $m'$ is infeasible.
\QED

\paragraph{Variant II.} 
The signature process is the following: 
\begin{description}
  \item[Public-key:] $[N,H(.)]$ 
  \item[Signed message:] $[m, F, R^3]$, where 
    $R$ is a secret random number, $S$ is a root of the equation $x^2=H(m) U$, where the padding factor $U$ is chosen as
    $U=\jacobi{H(m)}{p} \psi_1 + \jacobi{H(m)}{q} \psi_2$, and $F=RS$.
  \item[Verification:] Check whether $R^{12} H(m)^6=F^{12}$;  
     the signature is valid if and only if the equality holds.
\end{description}  

The algorithm works because $F^4= R^4 H(m)^2$, given that $U^2=1$. \\
For this scheme the verification cost is seven squares and three products, plus the evaluation of a hash function. 
It is possible to make different signatures of the same message by choosing different random numbers $R$. 

\begin{theorem}
   \label{forgV2}
The signature $[m, F, R^3]$ is weakly non-forgeable. 
It is weakly forgeable if $H(z)=z$ and strongly non-forgeable
 if $H(.)$ is a convenient hash function, in particular, if $H(z)=z(z+1)$.
\end{theorem}

\noindent
{\sc Proof}.    
Given $m'$, forgery is not possible because,  choosing w.l.o.g. $F'$, only a number $K$ such that $K H(m')^6=F'^{12}$ can be found, but not a fourth root of it. 
As above, weak forgeability in case of $H(z)=z$ follows from pseudo-homogeneity and strong non-forgeability from the hardness of inverting the hash function.
\QED

Note that using $R^2$ in the signature instead of $R^3$ would expose $S^2$ and therefore $U$, which would unveil the factorization of $N$ if $U$ is not $\pm 1$, but one of the other two roots of unity.

\subsection{Blind Rabin signature}
In principle, a blind Rabin signature is obtained as follows. Let $\mathbb A$ be the message author and $\mathbb B$ be the signer with public key $N$:

\begin{enumerate}
  \item $\mathbb A$ wants the message $m$ to be signed by $\mathbb B$ without disclosing the     message itself (or part of the message), then he chooses a random number $r$ and
    submits the disguised message $r^2 m$ to the signer.
  \item The signer $\mathbb B$ produces the signed message $[r^2 m, u, S]$, where $S$
   is a root of $x^2 =u r^2 m$, and $u$ is a random padding factor, and sends the
   signed message to $\mathbb A$.
  \item $\mathbb A$ receives the signed blind message $[r^2 m, u, S]$ and produces
   $[m, u, \frac{S}{r}]$, the signature for the original message. 
\end{enumerate}
This simple mechanism may be subject to forgery and to other kind of attacks, like for example the RSA blinding attack, which aims at using the blind signature protocol to decrypt messages that were encrypted using the public key of the signer. \\
Further, our Proposition \ref{blind} shows that the blind signer cannot use a strongly non-forgeable signature scheme; nevertheless, the open signed 
 message may be strongly non-forgeable. 
 
 Let $H(.)$ be a hash function used by the message author.  Consider the following process: \begin{description}
  \item[Public-key:] $[N,H(.)]$ 
  \item[Disguised message:] $r^2 H(m)$, where $m$ is the 
   original message to be signed, and $r$ is a random factor chosen by the author. 
   This message is submitted to the blind signer.
  \item[Blindly signed message:] $[r^2 H(m), F, R^3]$, where $F=R S$, with $R$ a random
   factor chosen by the signer, and $S$ a root of the quadratic equation $x^2=r^2 H(m) u$,
    the padding factor $u$ being defined as in Variant II.
  \item[Signed message:] $[m, \frac{F}{r^2}, \frac{R^3}{r^3}]$;  
  \item[Verification:] Check whether 
  $H(m)^6 \left(\frac{R^3}{r^3}\right)^4 =(\frac{F}{r^2})^{12}$;  
     the signature is valid if and only if the equality holds.
\end{description}  

\noindent 
The prime factors of the modulo $N$ are Blum primes, as we are using the scheme of Variant II. The verification cost is seven squares and three products, plus the evaluation of a hash function. \\
The signature of the original message is strongly non-forgeable, and the blind signature is not vulnerable to the RSA blinding attack as proved in the following theorem.

\begin{theorem}
   \label{forgblind}
The blind signature $[r^2 H(m), F, R^3]$, 
 is weakly non-forgeable and is not vulnerable to the RSA blinding attack.
The open signed message $[m, \frac{F}{r^2}, \frac{R^3}{r^3}]$ is strongly non-forgeable
 if $H(.)$ is a convenient hash function, in particular, if $H(m)=m(m+1)$.
\end{theorem}

\noindent
{\sc Proof}.
The blind signature $[r^2 H(m), F, R^3]$ is weakly forgeable as  
 $[t^2r^2 H(m), tF, R^3]$  
 for every $t \in \mathbb Z_N^*$, but to build a signature for a 
 given message $m'$ involves solving a quadratic equation which is unfeasible without knowing the factors of $N$, as already seen in discussing Variant II. 
The signature is not vulnerable to the RSA blinding attack because a square root
 of the message sent to the signer does not appear in the blind signature, as it is multiplied within $F$  by the random factor $R$ which is
 unknown to both author and attackers. \\
The author's signed message is taken as $[m, \frac{F}{r^2}, \frac{R^3}{r^3}]$
 with the blinding factor $r$ masking the random number $R$, for otherwise the signer
 may recognize the signed message by means of the random number $R^3$, thus breaking the anonymity. \\
Lastly, the signed message $[m, \frac{F}{r^2}, \frac{R^3}{r^3}]$ is strongly non-forgeable if $H(.)$ is a convenient hash function, in particular, if $H(m)=m(m+1)$, as seen in Theorem \ref{forgV2}.
\QED 

\section{Conclusions} 
In this paper we have presented several Rabin signature schemes and considered
 their resistances to forgery. We have also described blind Rabin signature schemes
  which are cryptographic primitives useful in protocols that guarantee the anonymity
  of the participants. In this kind of contexts, it is shown that the proposed
  schemes can be made resistant to the RSA blinding attack.

\section{Acknowledgments}
The research was supported in part by the Swiss National Science
Foundation under grant No. 132256. 


\end{document}